\documentclass[
]{ceurart}

\usepackage{graphicx}
\usepackage{booktabs}
\usepackage{eurosym}

\begin{document}

\copyrightyear{2021}
\copyrightclause{Copyright for this paper by its authors.
  Use permitted under Creative Commons License Attribution 4.0
  International (CC BY 4.0).}

\conference{AMPM'21: First Workshop in Agent-based Modeling \& Policy-Making, December 8, 2021, Vilnius, Lithuania}

\title{Empirically grounded agent--based policy evaluation of the adoption of sustainable lighting under the European Ecodesign Directive}

\author[1]{Gido H. Schoenmacker}[%
orcid=0000-0003-3946-928X,
email=gido@schoenmacker.nl,
]
\address[1]{Independent researcher, Amsterdam, the Netherlands}
  
\author[2]{Wander Jager}
\address[2]{University of Groningen, Groningen, the Netherlands}

\author[3]{Rineke Verbrugge}[%
orcid=0000-0003-3829-0106,
]
\address[3]{University of Groningen, Groningen, the Netherlands}

\begin{abstract}
Twelve years ago, the European Union began with the gradual phase-out of energy-inefficient incandescent light bulbs under the Ecodesign Directive. In this work, we implement an agent-based simulation to model the consumer behaviour in the EU lighting market with the goal to explain consumer behaviour and explore alternative policies. Agents are based on the Consumat II model, have individual preferences based on empirical market research, gather experience from past actions, and socially interact with each other in a dynamic environment. Our findings suggest that the adoption of energy--friendly lighting alternatives was hindered by a low level of consumer interest combined with high--enough levels of satisfaction about incandescent bulbs and that information campaigns can partially address this. These findings offer insight into both individual-level driving forces of behaviour and society--level outcomes in a niche market. With this, our work demonstrates the strengths of agent--based models for policy generation and evaluation.
\end{abstract}


\maketitle

\section{Introduction}

Twelve years ago, the European Union (EU) began with the gradual phase--out of energy--inefficient incandescent light bulbs under the Ecodesign Directive (2009/125/EC)~\cite{EU}. In 2019 it was estimated that this directive had reduced energy expenditure of household lamps by up to 60\%, saving the average EU family \euro$130$ annually~\cite{beuc2019}. Since this outcome benefits consumers, it might be expected that EU legislation was unnecessary for the adoption of energy--friendly lighting. The observed reality, however, was that household consumers had for years been hesitant to adopt more energy--friendly lighting options, prompting EU legislation~\cite{MILLS2010363}.

In this work, we implemented an agent--based simulation to model and explain the consumer behaviour in the EU lighting market. Because individual preferences and complex social interaction affect this behaviour, agent--based models are able to capture dynamics that would be difficult for traditional statistical models~\cite{castro2020review}. Our agents are based on the Consumat II model~\cite{Jager2012AnUC}: They have individual preferences based on consumer market research~\cite{kattenwinkel_2012}, gather experience from past actions, and socially interact with one another in a dynamic environment. We simulated multiple scenarios to examine consumer behaviour under different policies. In doing this, we aimed to answer two main questions: (i) Can we explain the reluctance of the consumer to switch to energy--friendly lighting? And (ii) Compared to banning household incandescent lighting, are there alternative policies that might have been equally successful?

\section{Methods}

\subsection{Model}

The Consumat II model -- Consumat for short -- is based on human psychological meta--theory and formulates drivers of complex behaviour in terms of needs, satisfaction, and uncertainty. It has been successfully applied to numerous, mostly environmentally--related, issues~\cite{JAGER2021133,Schaat2017,BRAVO2013225,kangur2017agent,moglia2018,van2019using}. When an action is required, a Consumat agent can engage in one of four cognitive strategies based on its position along two axes: its \textit{degree of satisfaction} and its \textit{degree of certainty}. The four strategies, in no particular order, are (i) repetition, (ii) imitation, (iii) deliberation, and (iv) social comparison.

Agents with a low degree of satisfaction are willing to expend effort to make changes, whereas high satisfaction leads to lower--effort strategies. Certainty relates to the degree of belief in expected outcomes when taking actions. A low level of certainty stimulates agents towards social strategies~\cite{bandura1962social,ellison1995word}, whereas high certainty encourages individually determined strategies~\cite{anand1995foundations,doi:10.1146/annurev-psych-122414-033417}. Agents with high satisfaction and certainty will engage in repetition, being satisfied with earlier actions and confident about results. High satisfaction combined with low certainty results in imitation, looking to peers for insurance in an uncertain environment. Agents with low satisfaction and high certainty engage in deliberation, trusting themselves to analyse the market and improve their situation. Finally, low satisfaction and certainty result in social comparison, where behaviour of peers is more closely examined and only copied if it is expected to increase satisfaction.

In our model, the behaviours are operationalised as follows. Repetition simply replaces a broken bulb with the same type of bulb. If a candidate bulb is no longer available, the agent will perform deliberation instead. Imitation selects a random but similar peer and replaces the broken bulb with a random available bulb. Peer difference is defined by $\| p_A - p_B \|_1$ where $p_N$ is the preference vector for agent $N$ (see below). If a randomly selected peer is not similar enough, the similarity requirement is loosened and a new peer is selected until a similar enough peer has been selected. Deliberation considers all available lamps and selects the one that will result in the highest satisfaction. Social comparison selects a peer in the same way as the imitation action and from its inventory selects the lamp that will result in the highest satisfaction.

\subsection{Model parameters}

To determine satisfaction, agents had individual preferences for lighting market--specific characteristics that were shown to significantly affect the decision process when purchasing light bulbs in market research by Kattenwinkel~\cite{kattenwinkel_2012}. In short, 97 Dutch individuals were questioned about their lamp purchase habits and considerations. After removing respondents with missing data, the resulting number of individuals was $87$. Examples of questions included the number of lamps respondents had in use and to which degree the opinion of social contacts affected lamp selection. The full list of questions is available in~\cite{schoenmacker_2014}.

Agent preferences were initialised based on archetypes from Kattenwinkel. In total, there are $87$ archetypes, where each archetype is an $11$--dimensional vector representing the number of lamps of an agents needs, tolerances for functional and colour requirements, focus on energy usage for financial and environmental reasons, social--mindedness and --agreeability, and baseline levels of experience/satisfaction with 3 different lighting types (incandescent, CFL, and light--emitting diode (LED)). In total $1000$ agents are instantiated by sampling uniformly from $(0.95v, 1.05v)$ for every characteristic $v$ from a single random archetype vector.

The lamp models and their properties in the models can be found in Supplemental Table~\ref{tab:s1}. Lamp lifetime is drawn from $\mathcal{N}(\ell,\,{\frac{\ell}{5}})$ where $\ell$ is the mean lifetime in months from Supplemental Table~\ref{tab:s1}. Every lamp is assumed in use for $6$ hours daily. Because no clear enough price progression is available, we assumed that LED lights become available in 2006 and then every year until 2020 decrease in price by 10\%. This means that their prices drop from \euro12.50--30.00 in 2006 to \euro3.20--7.60 in 2020. Since these may be influential assumptions, the price progression is varied by a random multiplication factor of 0.5--2 that governs LED pricing progress. Similarly, the energy efficiency of each LED bulb is assumed to grow with 5\% from 2007--2020 to a maximum of 99\% efficiency and is subject to a second, independent random factor 0.5--2 that governs LED innovation.

These changes to LED lamp properties occur in every scenario. Changes to the pricing and availability of incandescent lamps are scenario--dependent and are described in the ``Scenarios'' section below.

\subsection{Agent satisfaction}

Agent satisfaction describes how satisfied an agent is with a particular type of lamp. The following five lamp properties were found to significantly affect consumer behaviour by Kattenwinkel~\cite{kattenwinkel_2012} and thus included in the satisfaction function: colour discrepancy, energy efficiency, ramp--up time, and initial purchase price. These five properties were weighted by global relative importance from the same study and agent--specific properties, specifically tolerances for functional and colour requirements and focus on energy usage for financial and environmental reasons. More details including specific weighting values can be found in~\cite{schoenmacker_2014}.

\subsection{Scenarios}

\subsubsection{No regulation} In this scenario, there are no changes to incandescent lamp pricing or availability.

\subsubsection{Soft ban: regulation with mild effect} Because the effects of the Ecodesign Directive regulation are not clear--cut, we considered two possible scenarios. In this first one, the Ecodesign Directive results in an annual price increase of 10\% between 2013--2018 for incandescent bulbs, but does not affect availability. Like with the LED pricing and innovation, the incandescent price progression is varied by a third, independent random multiplication factor of 0.5--2. Because old batches and new bulbs for ``industrial use'' may still be sold, incandescent lamps may not not become wholly unavailable to the consumer.

\subsubsection{Hard ban: regulation with strong effect} In this second regulation scenario, incandescent lights become wholly unavailable to the consumer in 2015 after a 20\% price increase (with random multiplication factor) in the years 2012--2014. 

\subsubsection{Information campaign} This scenario is the same as the ``no regulation'' scenario with the change that in the year 2012, two agent properties are altered. The focus on energy usage for both financial and environmental reasons is increased by 50\%, simulating the possible effects of an information campaign that raises awareness of the benefits of lamp efficiency. As a result, agents in the model become more dissatisfied by energy inefficiency.

\subsubsection{Soft ban \& information campaign} This scenario combines the ``soft ban'' scenario with the ``information campaign`` scenario to investigate to which extent their effects are additive.\\

\noindent
Each scenario was run 50 times.

\subsection{Validation}

Due to the unavailability of numeric information about the progression of household lighting in the EU region, exact calibration and verification periods have not been assigned for our simulations. In the Conclusions we discuss plausibility of our results based on existing literature and technical reports.

\section{Results}

\begin{figure}
\begin{center}
\includegraphics[width=1\textwidth]{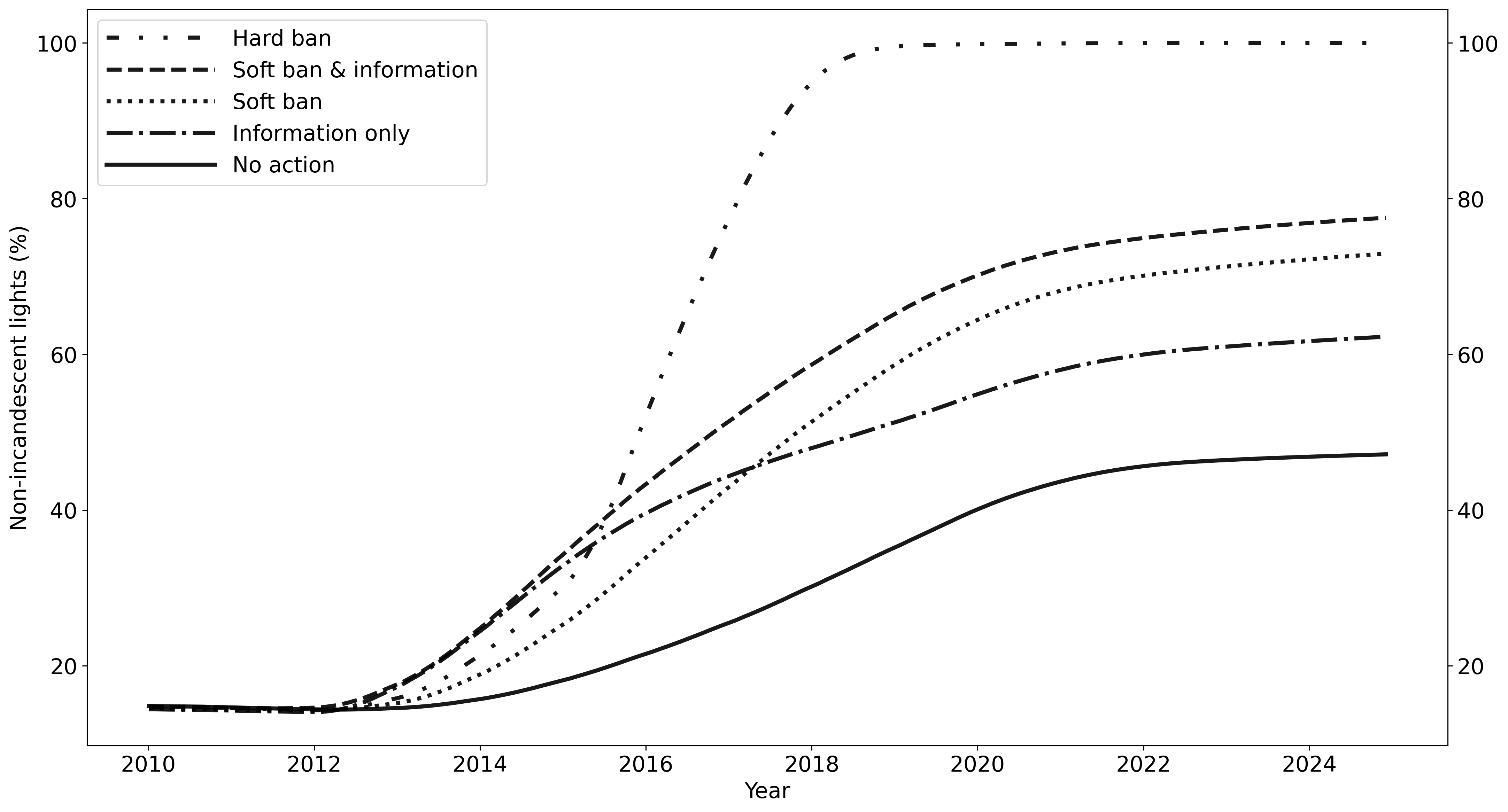}
\end{center}
\caption{Main simulation results showing the percentage of non--incandescent lamps in households over time. ``Non--incandescent'' is an umbrella term for halogen, CFL, LFL, HID, and LED lighting. Every line represents the mean results from 50 runs of a scenario. Similar plots with the standard deviation included are provided in the Supplement. The legend is sorted by efficacy, more effective measures at the top.}
\label{fig:1}
\end{figure}

The main results are shown in Figure~\ref{fig:1}, which contains the mean results for each scenario.

\subsection{No regulation} Firstly, we simulated the EU lighting market without any regulation. This resulted in slow adoption of energy--friendly alternatives: the incandescent light bulb remained dominant. Over 50\% of all lamps in consumer households remained incandescent bulbs by 2025. Supplemental Figure~\ref{fig:s5} shows the same plot as Figure~\ref{fig:1} with the standard deviation for this scenario included. When more energy--efficient LED lighting became affordable, a segment of the population that was previously engaged in deliberation and social comparison became satisfied and switched to repetition. This can be seen in Supplemental Figures~\ref{fig:s6}--\ref{fig:s9}, that contain the relative frequency of behaviours. However, the majority of the population remained unaffected. The main drivers for this behaviour appeared to be (i) the tendency of the Consumat to prioritise initial purchasing costs over total cost of ownership and (ii) a low level of interest in household lighting, leading to complacency with the functioning of incandescent lighting. 

\subsection{Soft ban} Secondly, we simulated two possible effects of the Ecodesign Directive policy. The ``soft ban'' consisted of a gradual price increase. This significantly increased LED adoption over the previous scenario: in 2025 around 75\% of lamps were non--incandescent. Supplemental Figure~\ref{fig:s3} shows the standard deviation for this scenario. Because incandescent bulbs were still available, and generally still the least expensive option at time of purchase, a segment of the Consumat population remained unaffected by market innovation. More and more Consumats who previously were unsatisfied with available options and engaged in high--effort strategies adopted the newer, more affordable LEDs and going forwards only engaged in repetition behaviour (Supplemental Figures~\ref{fig:s6},~\ref{fig:s8},~\ref{fig:s9}).

\subsection{Hard ban} Thirdly, we simulated a second possible effect of the Ecodesign Directive policy. In this scenario, incandescent lighting becomes unavailable to consumers. Trivially, the percentage of non--incandescent lighting quickly climbed to 100\% with low deviation (Supplemental Figure~\ref{fig:s1}). Initially there is a spike in deliberation behaviour forced by the impossibility of repetition (Supplemental Figure~\ref{fig:s8}). Next, many Consumats switch to social strategies, because the unavailability of their top choice resulted in higher uncertainty (Supplemental Figures~\ref{fig:s7} \& \ref{fig:s9}). Unhappiness with the remaining options means that repetition behaviour remains low (Supplemental Figure~\ref{fig:s6}) even long after the ban.

\subsection{Information campaign} Fourthly, we simulated an information--only policy that was aimed at informing the Consumat about energy--efficient lighting without restrictions on incandescent bulbs. This raised the energy--efficient lighting adoption by over 10 percentage points. The effects of the random variables governing LED pricing and innovation speed strongly affect adoption speed. If LED lighting quickly becomes more affordable, adoption reaches its peak around 2019. Slow price drops result in much lower peak adoption around 2022, not too different from the ``no regulations'' scenario (Supplemental Figure~\ref{fig:s4}). While consumer opinion was changed in this scenario, the main driving force of behaviour still appeared to be financial considerations.

\subsection{Soft ban \& information campaign} Lastly, we combined the ``soft ban'' and ``information campaign'' scenarios. While this scenario resulted in an increased adoption of around five percentage point over the ``soft ban'' only scenario, there was a large overlap between scenario outcomes as can be seen in Supplemental Figures~\ref{fig:s2} \& \ref{fig:s3}. Both the ``soft ban'' and ``information campaign'' scenarios appeared to reach a similar consumer audience, so that combining them does not result in a fully additive effect.

\section{Conclusions}

In this work, we implemented an agent--based simulation of the EU lighting market under different policies. Our main goals were to explain consumer behaviour and explore alternative policies. We found that our model was able to offer an explanation for the reluctance of the consumer to switch to energy--friendly lighting and that it could be used to investigate hypothetical policy scenarios. Our model suggested that the adoption of energy--friendly lighting alternatives was hindered by a low level of consumer interest combined with high--enough levels of satisfaction about incandescent bulbs. 

The most invasive scenario of making energy--inefficient bulbs unavailable was highly efficacious, because it forced the consumer to adapt its behaviour. A milder regulation that inflated prices of energy--inefficient bulbs, increased lighting adoption from below 50\% to around 75\% in 2025. An even less invasive option of an information campaign was less successful, increasing adoption to above 60\%. Combining milder regulation with an information campaign did not significantly increase adoption.

The authors are not aware of representative research measuring the effects of the Ecodesign Directive in the EU, either on the lighting market or the household lamp distribution. In~\cite{beuc2019}, effects of the Ecodesign Directive are estimated based on different hypothetical scenarios. Similarly, the Model for European Light Sources Analysis~\cite{MELISA} presents numbers up until 2013. Koretsky~\cite{KORETSKY2021102310} notes that in 2020, incandescent bulbs are still available to purchase online. The most complete source on the evolution of the lighting market may be Zissis et al.~\cite{zissis2021update} showing that in 2019, fluorescent and LED sales each made up about half of \textgreater$90$\% of global sales, suggesting a \textless$10$\% market share for incandescent lighting. This would mean that actual non--incandescent penetration lies between our hard and soft ban scenarios. This makes sense insofar that the actual market results of the Ecodesign Directive also appear to lie in between these two scenarios: incandescent bulbs are still available for purchase online and in hardware stores, but not readily available (e.g. in supermarkets). Our scenarios did not consider the effects of limited availability.

From the four possible behaviours in our model, repetition was by far the most common (Supplemental Figure~\ref{fig:s6}). Making non--disruptive changes to the market or consumer opinion did increase adoption rate, but it failed to reach consumers who were set in their ways. Our findings agree with literature in that, while social interactions are known to significantly affect customer repeat behaviour (e.g.~\cite{doi:10.1509/jmr.13.0482}), consumer habits are hard to break without direct disruption~\cite{doi:10.1509/jppm.25.1.90}. Financial savings, increased availability, more natural colouring, and environmental concerns are mentioned as leading factors in LED adoption~\cite{leelakulthanit2014factors,cowan2011understanding,dangol2015user,Cowan2013} and also implicated by our model.

Randomness occurs in our model through consumer preference instantiation, peer selection in social behaviour, and three random factors of $0.5$--$2$ governing LED pricing decrease, incandescent pricing increase, and LED innovation progression (increased life span and colour temperature of LEDs). Sensitivity analysis showed that the largest variation as seen in Figures~\ref{fig:s1}--\ref{fig:s5} occurs through the interaction of the two factors governing incandescent and LED pricing. Even though our agents consider other factors, initial purchasing price remains a significant consideration for many. This means that a tipping point is reached as soon as incandescent bulbs are no longer the least expensive option in stores, the occurrence and timing of which in our model is determined by the two random factors governing price progression.

With our model, we showed an application of agent--based models in explaining consumer behaviour and testing policies to affect this behaviour. A body of related work using agent--based methods specifically for climate--relevant behaviours exists~\cite{castro2020review,JAGER2021133}. Closely related to our application, Hicks et al.~\cite{https://doi.org/10.1111/jiec.12281} studied the innovation diffusion of energy saving lighting focusing on information and perception, concluding that increased usage may (partially) offset energy savings. Relatedly, Muelder and Filatova~\cite{muelder2018} investigated different formalisations of social theories in energy consumption, specifically solar investments. Buskens~\cite{BUSKENS2020485} summarised the sociological background of innovation diffusion in social networks and concluded that close social circles are especially important to establish trust in new products.

Our work is limited by a number of assumptions about the development of the lighting market and the effects of information campaigns, as well as the abstraction of consumer behaviour into four strategies. We measured the effects of our assumptions on agent preferences, pricing progression, and technical innovation by including independent random variables that halved to doubled our projections, exploring a wide range of possible progressions. Our model environment and agent preferences are realistic by virtue of being based on empirical market research. The Consumat model itself, while necessarily being a simplified abstraction, is strongly grounded in psychological theory and exhibits complex behaviour at macro-levels~\cite{Jager2012AnUC}. This combination allowed us to generate specific testable hypotheses about consumer behaviour.

In conclusion, our agent--based model of the lighting market was able to explain consumer behaviour and evaluate counterfactual policies. Our findings offer insight into both individual--level driving forces of behaviour and society--level outcomes in a niche market. With this, our work demonstrates the strengths of agent--based models for policy generation and evaluation.

\bibliography{refs}

\clearpage

\appendix

\section*{Supplement}

\renewcommand{\thetable}{S\arabic{table}}
\renewcommand{\thefigure}{S\arabic{figure}}

\begin{table}[!htb]
\caption{Initial settings for the lamps available to the model.  The high initial prices for LED lighting are based on 2006 prices, when LED was just becoming available to consumers. The ``Available'' column represents availability at the beginning of the simulation and changes over time (see main text).}
\label{tab:s1}
\begin{center}
\begin{tabular*}{\textwidth}{c @{\extracolsep{\fill}} lcccccc}
\toprule
Type & Price & Efficiency & Colour & Ramp--up & Lifetime & Available \\
     & \euro & \% & \% & Sec. & Months & Y/N \\
\midrule
LED             & 30.00     & 63              & 10              &  1              & 125         & N  \\
LED             & 25.00     & 60              & 10              &  1              & 167         & N  \\
LED             & 20.00     & 60              & 10              &  1              & 208         & N  \\
LED             & 15.00     & 60              & 15              &  1              & 167         & N  \\
LED             & 12.50     & 60              & 10              &  2              & 208         & N  \\
CFL             &  9.30     & 80              & 30              & 80              &  83         & Y  \\
CFL             &  8.40     & 90              & 15              & 80              &  83         & Y  \\
CFL             &  7.80     & 90              & 15              & 40              & 100         & Y  \\
CFL             &  7.80     & 90              & 15              & 40              &  83         & Y  \\
CFL             &  7.00     & 90              & 15              & 40              &  83         & Y  \\
CFL             &  5.00     & 70              & 15              &  1              &  17         & Y  \\
CFL             &  3.20     & 70              & 15              &  1              &   8         & Y  \\
CFL             &  3.20     & 60              & 15              &  1              &  17         & Y  \\
CFL             &  2.50     & 60              & 15              &  1              &  17         & Y  \\
Incandescent    &  3.00     & 30              &  5              &  1              &  17         & Y  \\
Incandescent    &  2.70     & 50              &  5              &  1              &   8         & Y  \\
Incandescent    &  1.80     & 50              &  5              &  1              &   8         & Y  \\
Incandescent    &  1.80     & 40              &  5              &  1              &   8         & Y  \\
Incandescent    &  1.40     & 50              &  5              &  1              &   8         & Y  \\
\bottomrule
\end{tabular*}
\end{center}
\end{table}

\begin{figure}
\begin{center}
\includegraphics[width=1\textwidth]{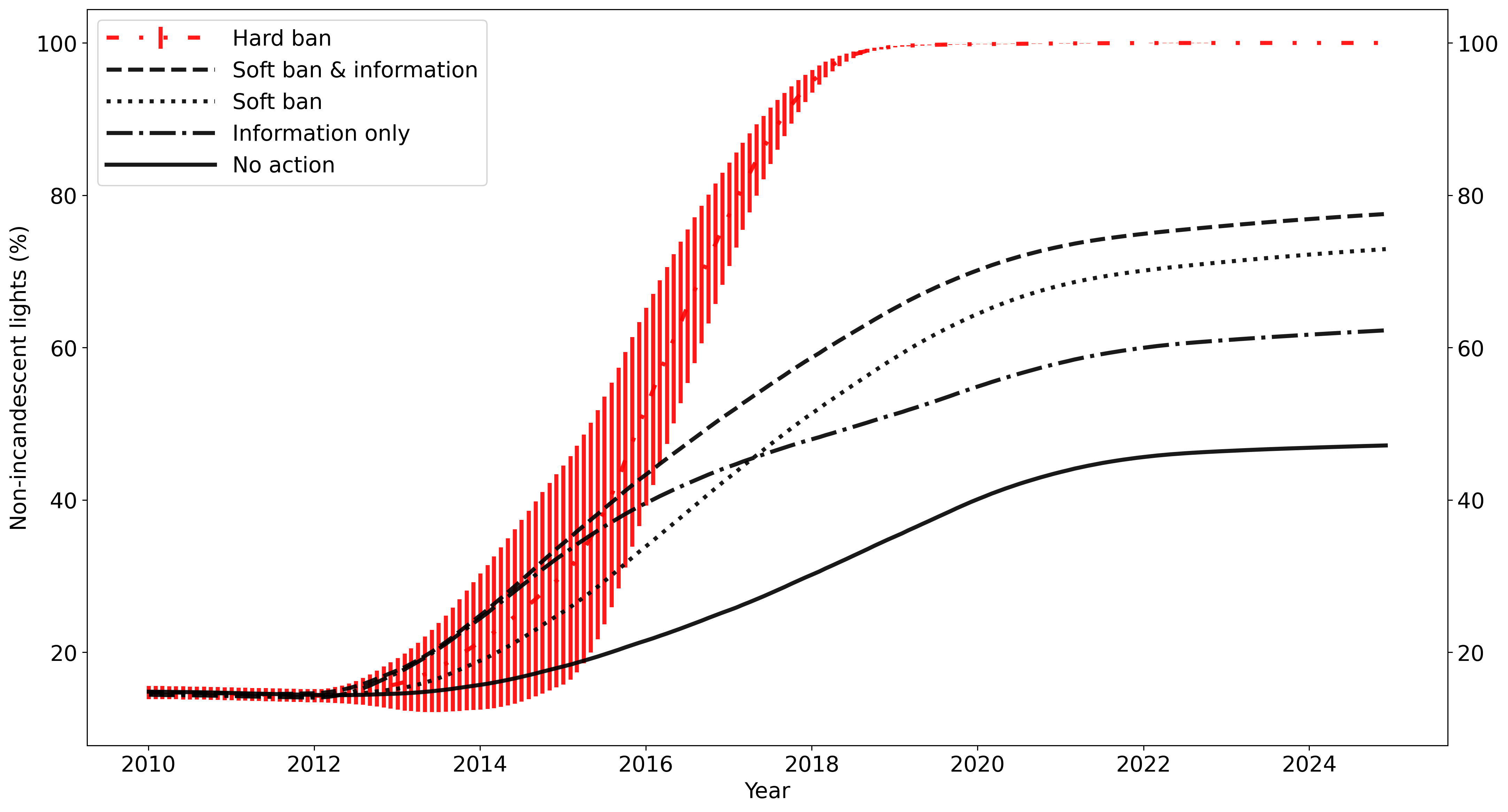}
\end{center}
\caption{Main simulation results showing the percentage of non--incandescent lamps in households over time. ``Non--incandescent'' is an umbrella term for halogen, CFL, LFL, HID, and LED lighting. Every line represents the mean results from 50 runs of a scenario. Standard deviation of the ``hard ban'' scenario is included in red.}
\label{fig:s1}
\end{figure}

\begin{figure}
\begin{center}
\includegraphics[width=1\textwidth]{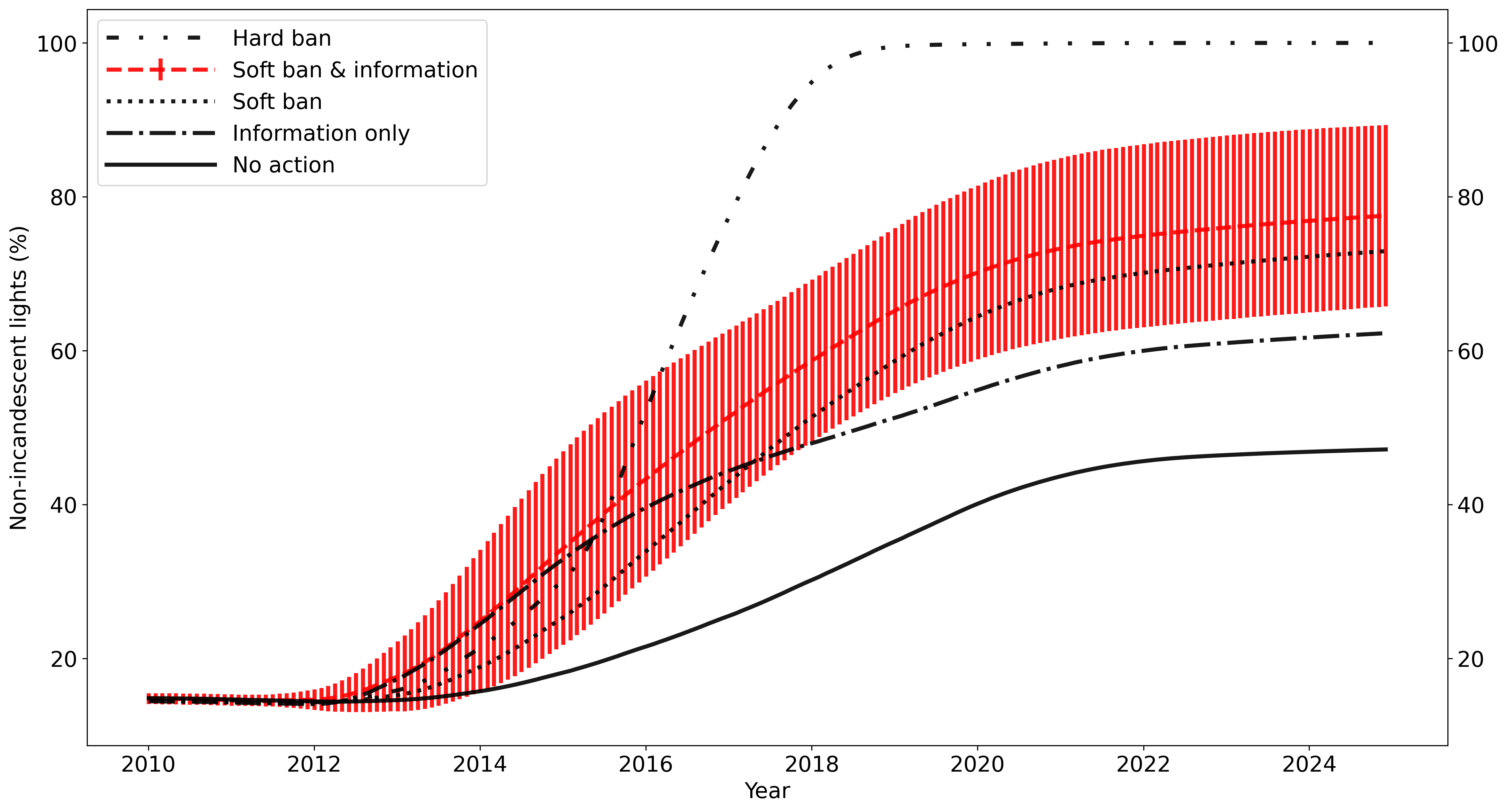}
\end{center}
\caption{Simulation results showing the percentage of non--incandescent lamps in households over time. ``Non--incandescent'' is an umbrella term for halogen, CFL, LFL, HID, and LED lighting. Every line represents the mean results from 50 runs of a scenario. Standard deviation of the ``soft ban \& information campaign'' scenario is included in red.}
\label{fig:s2}
\end{figure}

\begin{figure}
\begin{center}
\includegraphics[width=1\textwidth]{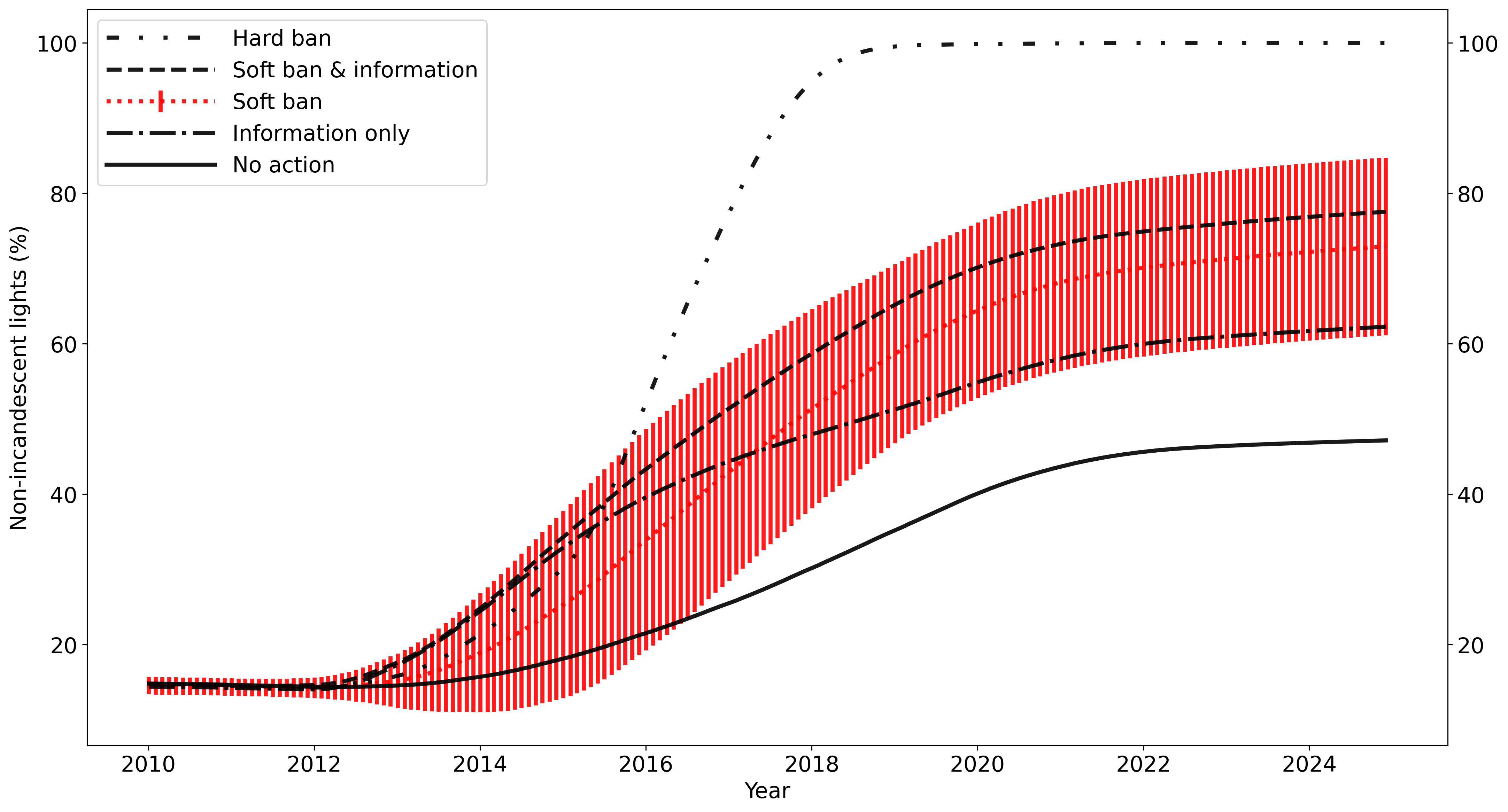}
\end{center}
\caption{Simulation results showing the percentage of non--incandescent lamps in households over time. ``Non--incandescent'' is an umbrella term for halogen, CFL, LFL, HID, and LED lighting. Every line represents the mean results from 50 runs of a scenario. Standard deviation of the ``soft ban'' scenario is included in red.}
\label{fig:s3}
\end{figure}

\begin{figure}
\begin{center}
\includegraphics[width=1\textwidth]{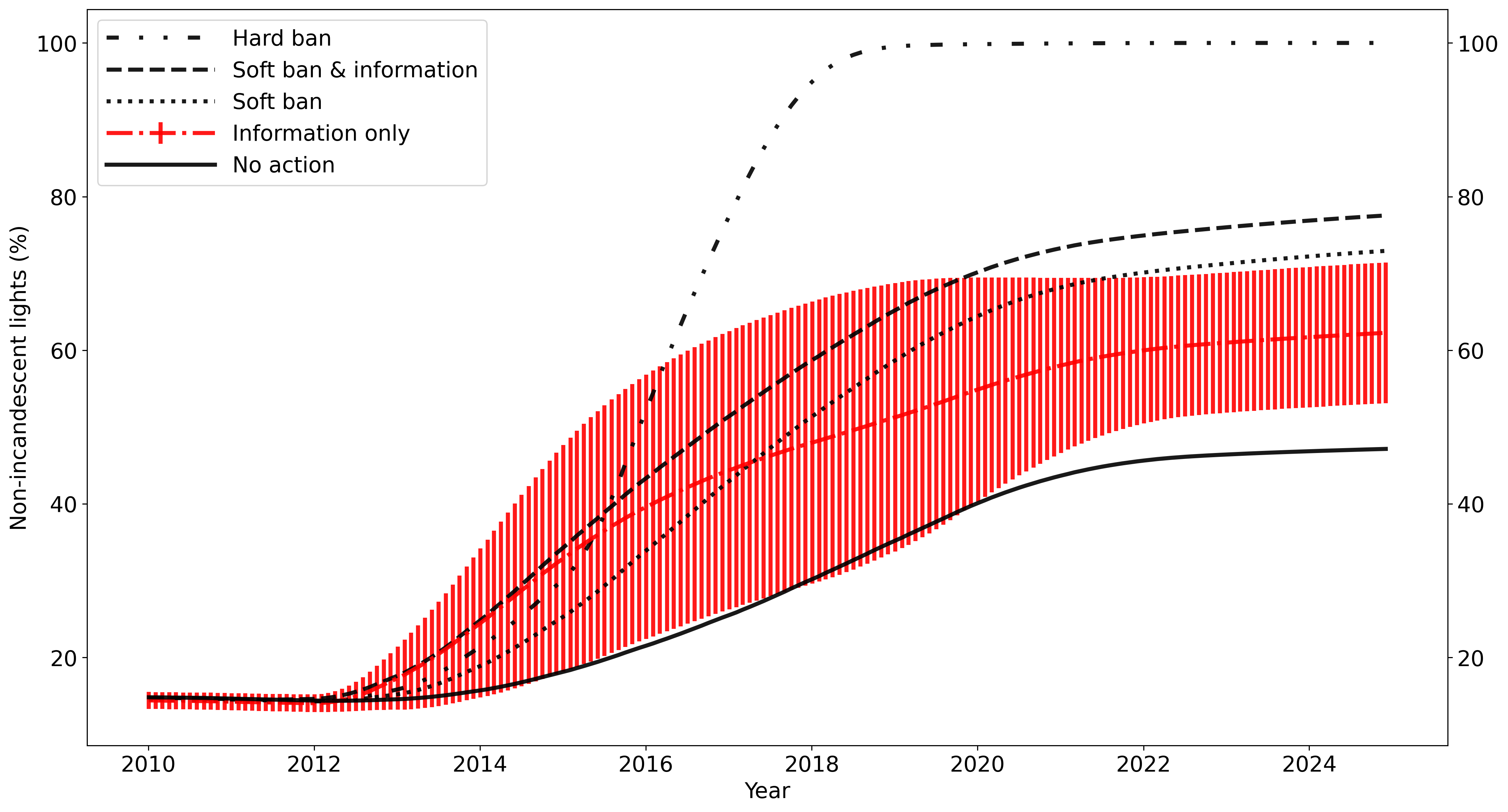}
\end{center}
\caption{Simulation results showing the percentage of non--incandescent lamps in households over time. ``Non--incandescent'' is an umbrella term for halogen, CFL, LFL, HID, and LED lighting. Every line represents the mean results from 50 runs of a scenario. Standard deviation of the ``information campaign'' scenario is included in red.}
\label{fig:s4}
\end{figure}

\begin{figure}
\begin{center}
\includegraphics[width=1\textwidth]{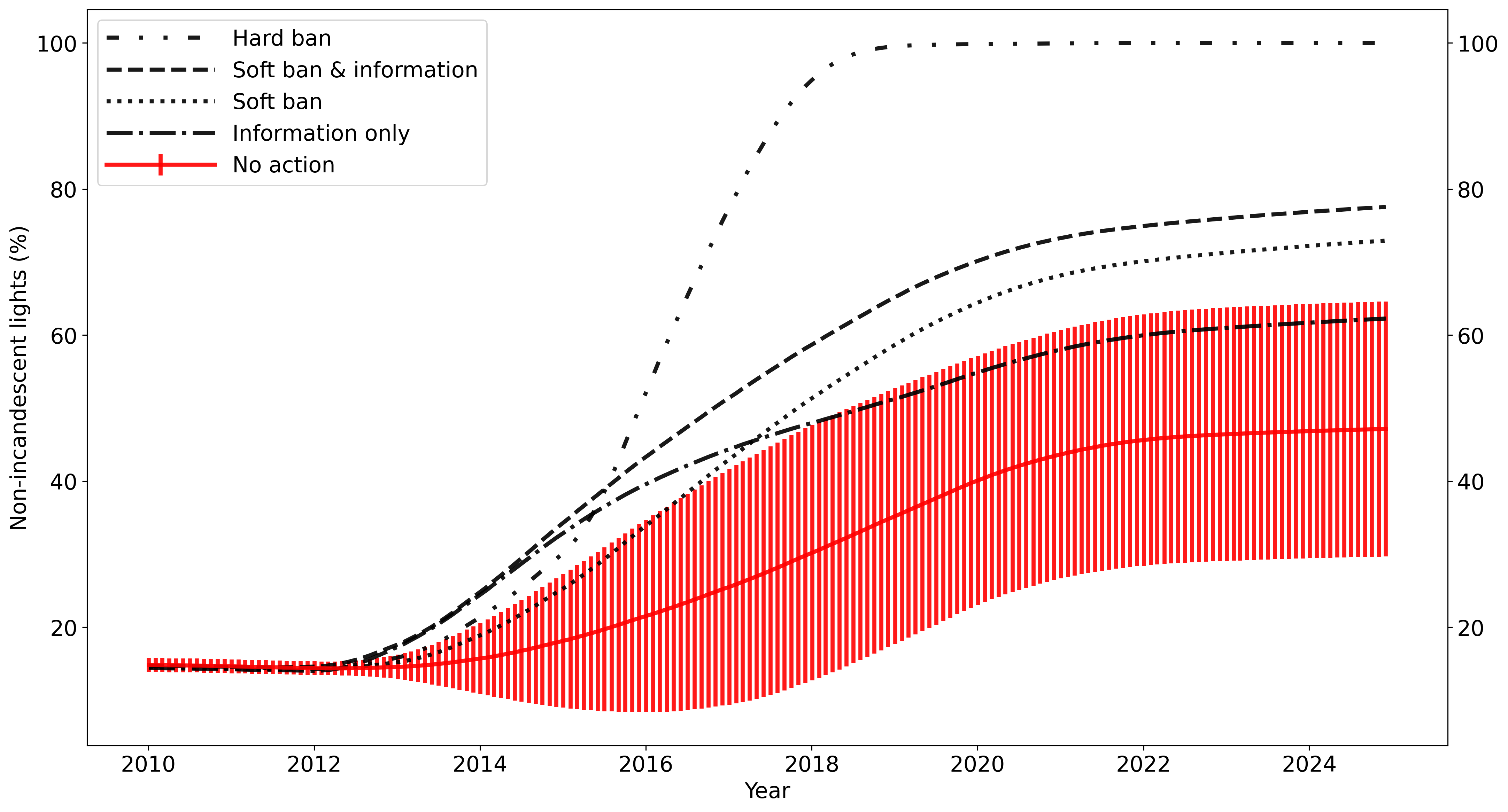}
\end{center}
\caption{Simulation results showing the percentage of non--incandescent lamps in households over time. ``Non--incandescent'' is an umbrella term for halogen, CFL, LFL, HID, and LED lighting. Every line represents the mean results from 50 runs of a scenario. Standard deviation of the ``no regulation'' scenario is included in red.}
\label{fig:s5}
\end{figure}

\begin{figure}
\begin{center}
\includegraphics[width=1\textwidth]{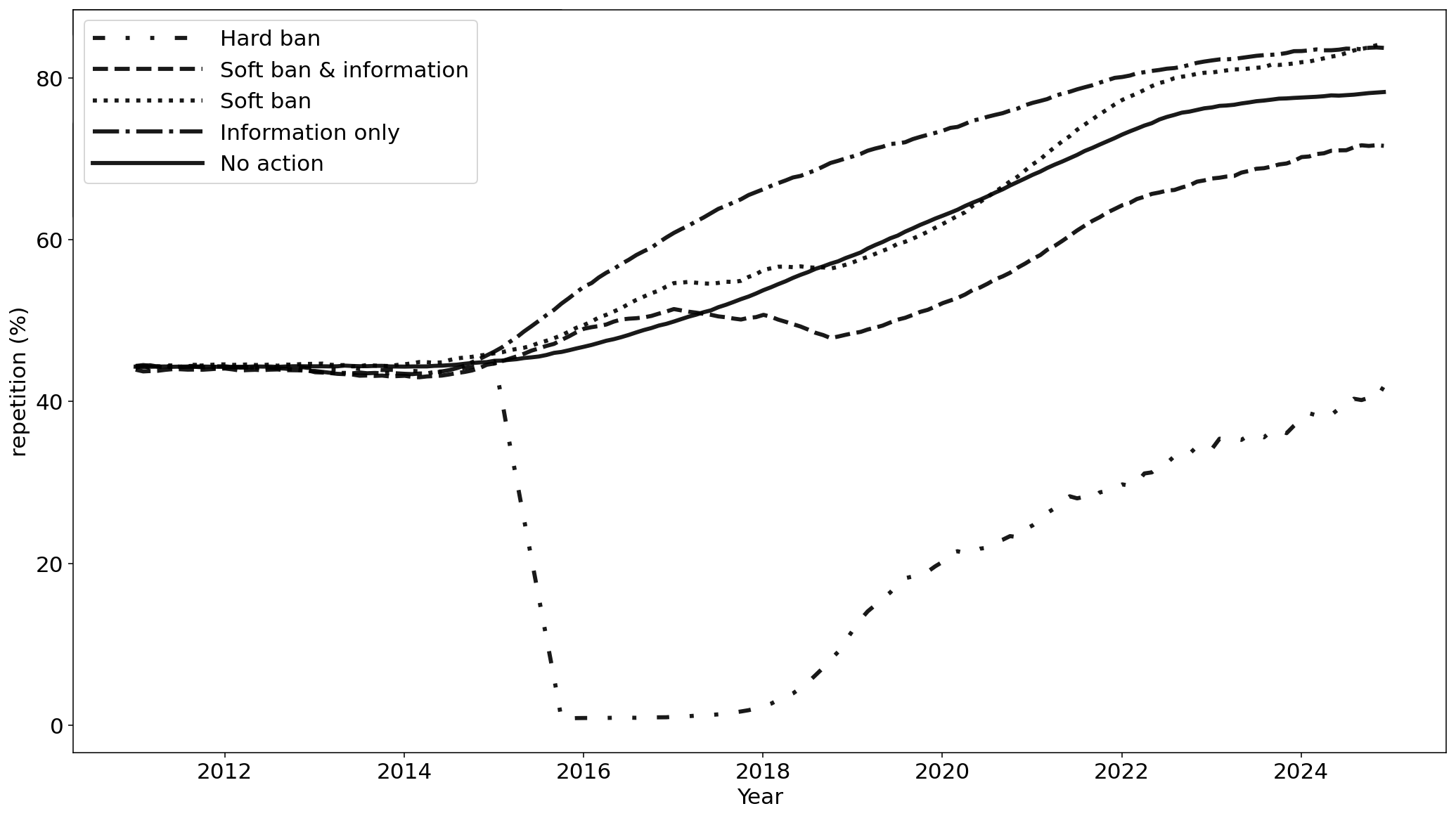}
\end{center}
\caption{Simulation results showing percentage of repetition behaviour over time.}
\label{fig:s6}
\end{figure}

\begin{figure}
\begin{center}
\includegraphics[width=1\textwidth]{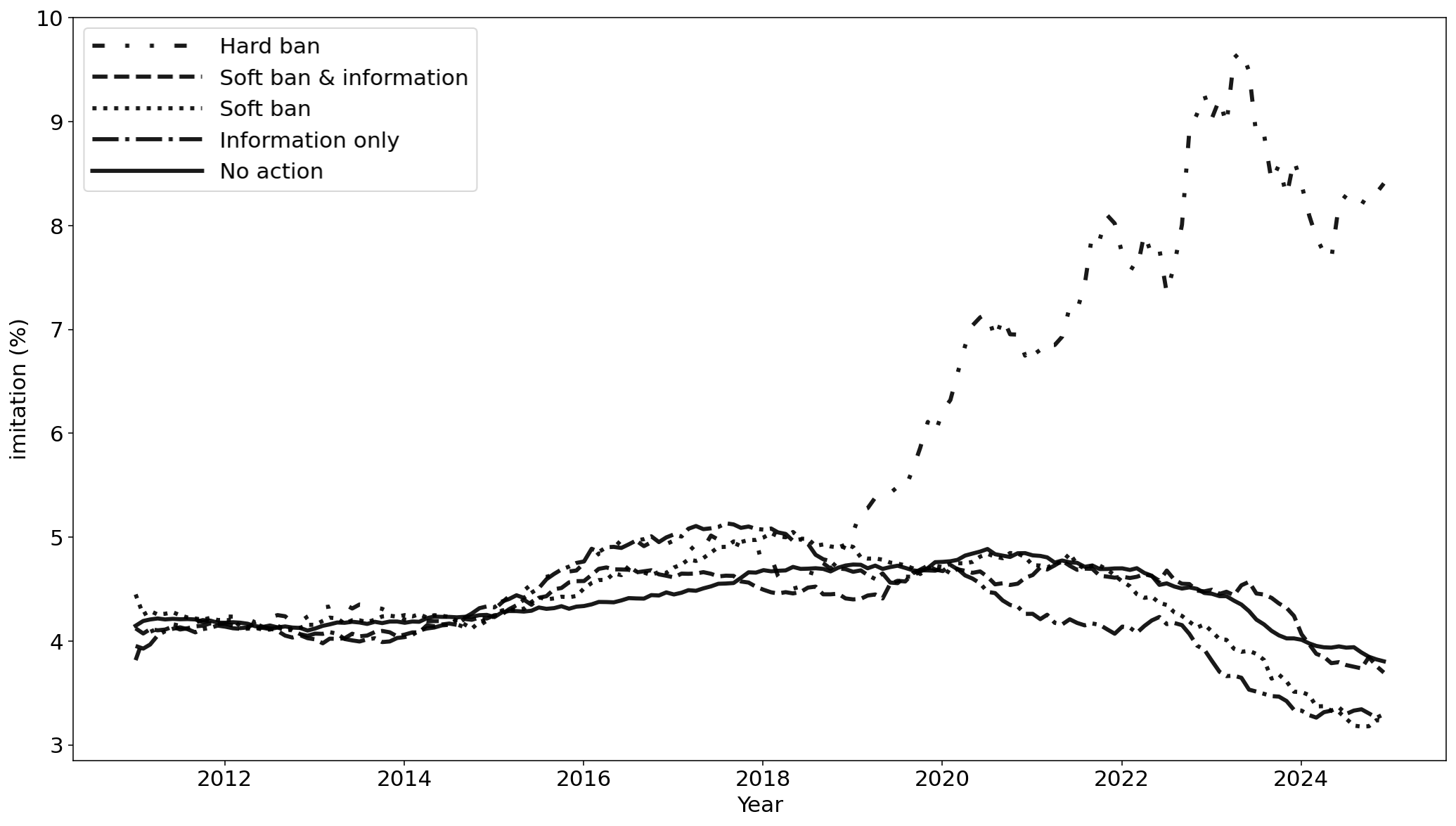}
\end{center}
\caption{Simulation results showing percentage of imitation behaviour over time.}
\label{fig:s7}
\end{figure}

\begin{figure}
\begin{center}
\includegraphics[width=1\textwidth]{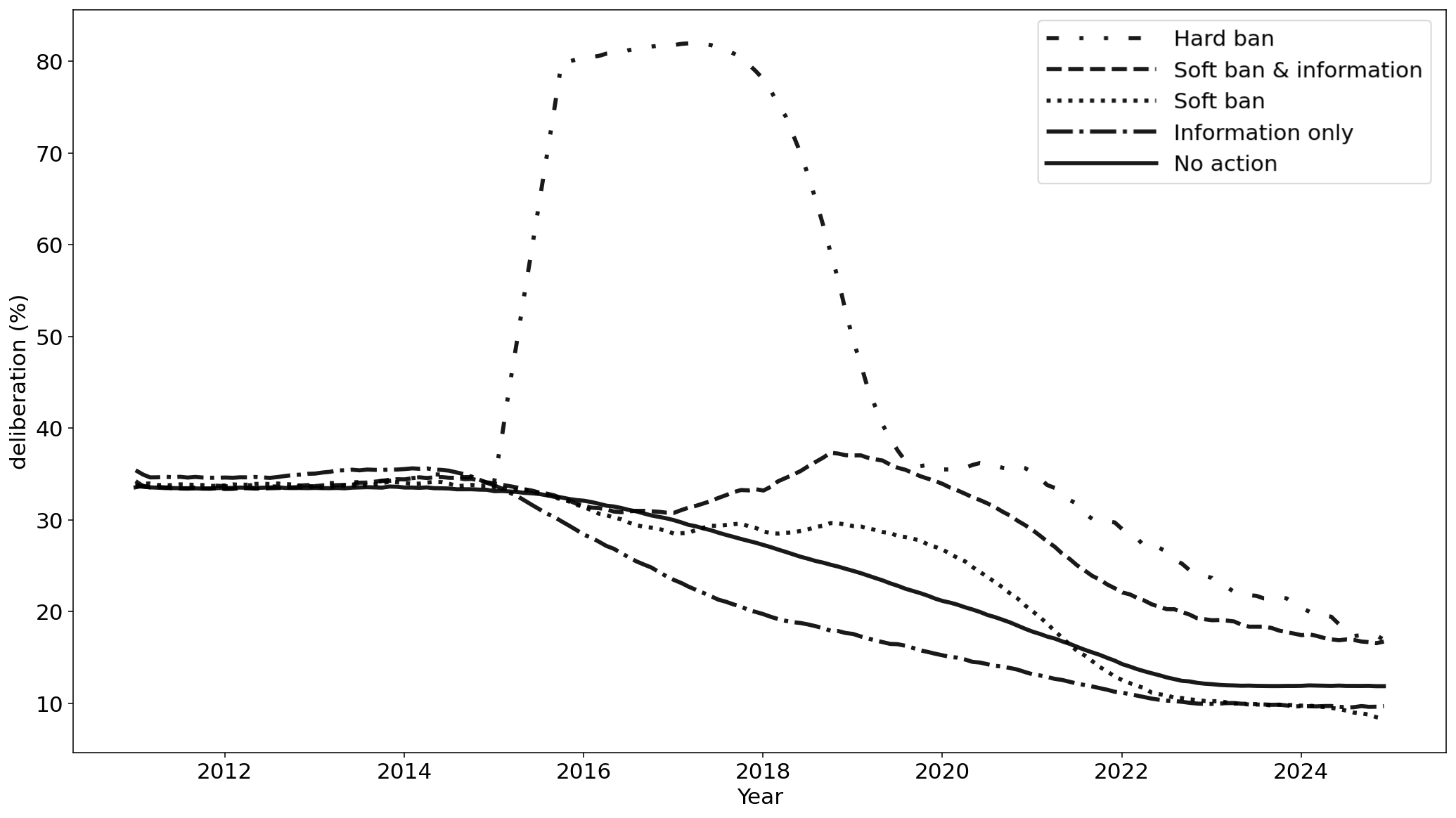}
\end{center}
\caption{Simulation results showing percentage of deliberation behaviour over time.}
\label{fig:s8}
\end{figure}

\begin{figure}
\begin{center}
\includegraphics[width=1\textwidth]{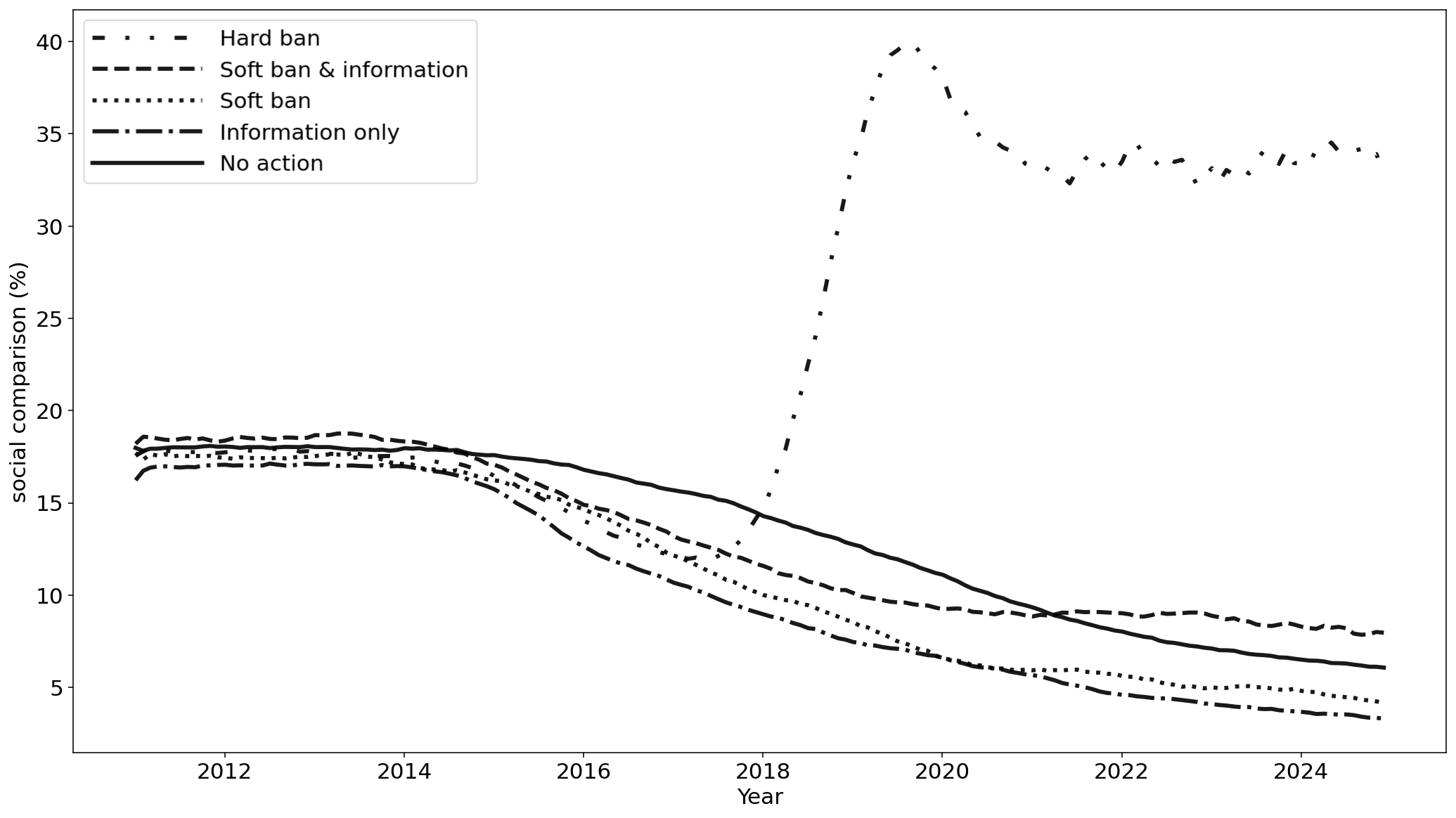}
\end{center}
\caption{Simulation results showing percentage of social comparison behaviour over time.}
\label{fig:s9}
\end{figure}

\end{document}